# Valley-Selective Linear Dichroism and Excitonic Effects in Lieb-Lattice Altermagnets

Haonan Wang[1], Xilong Xu[1], Du Li[1] and Li Yang[1,2*]

1. Department of Physics, Washington University, St. Louis, MO, 63130, USA

2. Department of Physics and Institute of Materials Science and Engineering, Washington University, St. Louis, MO, 63130, USA

*lyang@physics.wustl.edu

**Abstract**

Altermagnets have recently been recognized as a distinct class of magnetic materials characterized by alternative spin-split electronic structures without net magnetization. Despite intensive studies on their single-particle spintronic and valleytronic properties, many-electron interactions and optical responses of altermagnets remain less explored. In this work, we employ many-body perturbation theory to investigate excited states and their strain tunability. Using monolayer $Mn_2WS_4$ as a representative candidate, we uncover a novel spin–valley-dependent excitonic selection rule in two-dimensional altermagnetic Lieb-lattices. In addition to strongly bound excitons, we find that linearly polarized light selectively excites valley-spin-polarized excitons. Moreover, due to the interplay between altermagnetic spin symmetry and electronic orbital character, we predict that applying uniaxial strain can lift valley degeneracy and enable the selective excitation of spin-polarized excitons—an effect not achievable in previously studied transition-metal dichalcogenides. These spin-valley-locked excitonic states and their strain tunability offer a robust mechanism for four-fold symmetric altermagnets to encode, store, and read valley/spin information.

**Introduction**

Altermagnets have recently emerged as a promising platform for exploring unconventional quantum phenomena [1–3]. A hallmark of altermagnets is their unique spin symmetry, which gives rise to alternative spin-split electronic structures despite the absence of both net macroscopic magnetization and spin–orbit coupling (SOC). This unconventional magnetic ordering establishes altermagnets as a third fundamental magnetic phase, distinct from conventional ferromagnetism and antiferromagnetism. To date a growing number of materials have been experimentally identified or theoretically predicted to exhibit the altermagnetic order, including MnTe [4–7], CrSb [8–10], $RuO_2$ [11–14], $V_2Se_2O$ family [15–18], among others. Numerous theory works have proposed abundant novel quantum phenomena, such as anomalous Hall effect [14,19–23], piezomagnetism and alter-piezoresponses [15,18,24], multiferroics [25–30], and spin-current generation [15,31] and so on [32–36].

Among many exotic properties of altermagnets, a new form of spin-valley locking (SVL) driven by intrinsic altermagnetic spin symmetry has been revealed [15,37,63], offering exciting opportunities for manipulating spin and valley degrees of freedom. Unlike SVL in transition-metal dichalcogenides (TMDs), this effect does not rely on SOC. As a result, altermagnets are expected to exhibit a longer spin depolarization time, which is critical attribute in spintronic applications [38]. Consequently, altermagnets hold significant promise for future spintronic and valleytronic applications [39]. However, in contrast to time-reversal-symmetric materials such as graphene and TMDs—where the interaction between light and valley degrees of freedom has been extensively studied [30, 40–42], the influence of SVL on the optical properties of altermagnets has not been fully uncovered. Furthermore, beyond single-particle physics and model predictions of excitonic effects on altermagnets [63], first-principles calculated many-body effects, which are known to be greatly enhanced in low-dimensional systems and determine their optical responses, remain largely unexplored in the context of altermagnetic materials.

In this work, we investigate the valley-resolved optical and excitonic properties of two-dimensional (2D) Lieb-lattice altermagnets with the $S_4T$ symmetry, using monolayer $Mn_2WS_4$ as an example. Employing first-principles many-body perturbation theory (MBPT), we reveal a novel valley-selective linear dichroism arising from the interplay between altermagnetic spin symmetry and orbital character. Our results uncover a spin-valley-dependent optical selection rule, where linearly polarized light can selectively excite excitons in different valleys with opposite spins, resulting in a series of doubly degenerate exciton states located below the quasiparticle gap. These strongly bound excitons are found to be fully spin-polarized and highly anisotropic in both wavefunctions and optical polarization. Furthermore, we demonstrate that applying uniaxial strain can lift the valley degeneracy, enabling selective and directional tuning of excitonic energies and optical responses. Together, these findings highlight the rich exciton-based valleytronic and spintronic potential of four-fold symmetric altermagnets and lay the groundwork for novel spin-optoelectronic applications.

**Computational Details**

The ground-state properties of $Mn_2WS_4$ were calculated using density functional theory (DFT) within the generalized gradient approximation (GGA) with the Perdew-Burke-Ernzerhof (PBE) exchange-correlation functional [43] as implemented in the Quantum ESPRESSO package [44,45]. Norm-conserving pseudopotentials [46] were used for all elements, including semi-core states for Mn atoms. The plane-wave cutoff energy was set to 65 Ry. A vacuum spacing of 18 Å was used to avoid spurious interactions between periodic images in monolayer calculations. SOC is not included because its negligible influence on the optical responses (see supplementary information). To account for the on-site Coulomb repulsion of Mn $3d$ electron, the LDA+U method [47] is applied with Hubbard $U = 3.0$ eV. MBPT calculations were performed using the BerkeleyGW code [48]. Quasiparticle energies were computed using the single-shot $G_0W_0$ approximation with a general plasmon-pole model [49]. Excitonic effects and optical spectra were calculated by solving the Bethe-Salpeter Equation (BSE) [50], including Coulomb truncation for monolayer geometry. A $12 \times 12 \times 1$ $k$-grid was used for self-energy and dielectric calculations and interpolated to a $48 \times 48 \times 1$ grid for solving the BSE for achieving the converged results. Six valence and six conduction bands were included to ensure convergence of excitonic states and optical spectra below 3.5 eV.

**Results and Discussion**

The ternary transition-metal chalcogenides, $M_2WX_4$ (M=Mn, Fe and X=S, Se), crystallize in the tetragonal space group $P\bar{4}2m$ (No. 111), which belongs to the $D_{2d}$ point group. The structure consists of a three-sublayer atomic configuration in which a central M–W atomic plane is sandwiched between two chalcogen layers, as illustrated in Fig. 1(a). Each unit cell contains two M sublattices separated by a single tungsten atom, collectively forming a Lieb lattice. Notably, the ternary transition-metal chalcogenides of $M_2WX_4$ (M = Fe, Cu and X = S, Se) have been previously reported [24,32,51–55], including experimentally synthesized layered nonmagnetic compounds.

Our first-principles calculations indicate that the magnetic ground states of monolayer $M_2WX_4$ (M = Mn, Fe, and X = S, Se) are Néel-type antiferromagnet, as shown in Fig. 1(a). In this work, we focus on Mn$_2$WS$_4$ as a representative example to demonstrate the excitonic effects. First-principles calculations yield an in-plane lattice constant of 5.443 Å, and each Mn atom carries a local magnetic moment around 4.2 $\mu_B$. The overlap of wavefunctions between neighboring manganese sublattices Mn$_1$, Mn$_2$, and sulfur ions facilitates virtual hopping through Mn–S–Mn superexchange pathways, resulting in a robust antiferromagnetic coupling, as depicted in Fig. 1(b). Thus, the spin-up and spin-down sublattices are transformed via the spin-group symmetry $[C_2\|S_{4z}]$, where $C_2$ represents a spin-flip operation and $S_{4z}$ denotes a rotoinversion operation that breaks the spatial inversion symmetry. Based on this symmetry classification, monolayer Mn$_2$WS$_4$ belongs to $d$-wave $S_4T$-symmetric altermagnetic materials, possessing $C_{2x}T$, $C_{2y}T$, $S_{4z}T$, and $M_{xy}$ symmetry. In Fig. 1(c), we present the GW-calculated band structures, and the quasiparticle gap is approximately 1.7 eV, which is about 1.2 eV larger than the DFT value. As expected, the band structure exhibits prominent spin splitting along the Γ-X-M-Y-Γ high-symmetry path, while the energy bands along Γ-X-M are degenerate with those along Γ-Y-M with opposite spin, as shown in Fig. 1(c). Remarkably, the spin splittings of both conduction and valence bands at the X and Y valleys reach several hundred meV, significantly exceeding those observed in many known altermagnets [2] and representing a desirable feature for spintronic applications.

In particular, the band structures of monolayer Mn$_2$WS$_4$ exhibits SVL as well. As shown in Fig. 1(c), both the topmost valence band and bottommost conduction band are fully spin-polarized: spin-up at the X valley and spin-down at the Y valley, respectively. Such SVL strongly indicates the potential optical manipulation over both spin and valley degrees of freedom.

We proceed to investigate the optical properties of monolayer Mn$_2$WS$_4$. The single-particle optical absorption spectrum is shown in Fig. 2(a) for the incident light polarized along the axial $x$-direction. As expected, the absorption edge corresponds closely to the quasiparticle gap (~1.7 eV). After analyzing the origin of the absorption peak, we find a nearly perfect spin-valley-selective linear dichroism. As shown in Figs. 2(a) and 2(b), the absorption profiles are identical for $x$-polarized and $y$-polarized incident lights. However, their interband transition origins in reciprocal space are locked to different valleys: the optical absorption under $x$- ($y$-) polarization originates mainly from interband transitions near the X (Y) valley, as highlighted in the insets of Figs. 2(a) and 2(b).

To quantitatively characterize the spin-valley-dependent linear dichroism, we define the degree of linear optical polarization, $\eta(\mathbf{k})$, following the convention of circular valley polarization commonly used in transition metal dichalcogenides (TMDs) [40], as follows:

$$\eta(\mathbf{k}) = \frac{|P_x^{cv}(\mathbf{k})|^2 - |P_y^{cv}(\mathbf{k})|^2}{|P_x^{cv}(\mathbf{k})|^2 + |P_y^{cv}(\mathbf{k})|^2}, \quad (1)$$

where $\mathbf{P}^{cv}(\mathbf{k}) = \langle \psi_{c\mathbf{k}} | \mathbf{p} | \psi_{v\mathbf{k}} \rangle$ is the interband transition matrix element between the topmost valence and bottommost conduction bands, and $P_x^{cv}(\mathbf{k})$ and $P_y^{cv}(\mathbf{k})$ correspond to transitions under linearly polarized light along $x$- and $y$-directions, respectively. As shown in Fig. 2(c), the momentum-resolved map of $\eta(\mathbf{k})$ clearly reveals four-fold alternating regions of positive (blue) and negative (red) values centered around the X and Y valleys. Under linearly polarized excitation, $\eta(\mathbf{k}) = \pm 1$ at these valleys, confirming that the interband transitions couple exclusively to orthogonal linear polarizations—parallel and perpendicular to the crystalline $x$-axis, respectively. This behavior manifests as valley-selective linear dichroism, a novel symmetry-driven optical phenomenon. Analogous to the well-known valley-selective circular dichroism observed in monolayer TMDs [40], where the K and K′ valleys selectively absorb left- and right-handed circularly polarized light, valley-selective linear dichroism enables the distinction and control of valleys using linearly polarized photons. This provides a new axial degree of freedom for valleytronic and spintronic applications.

The novel optical selection rules essentially arise from the symmetry properties of the involved atomic orbitals, which are illustrated in Fig. 2(a) and 2(b). At the X and Y valleys in the Brillouin zone of space group $P\bar{4}2m$, the little group is C$_{2v}$. Orbital-projected analysis shows that the topmost valence bands at the X and Y valleys predominantly originate from the $d_{xz}$ and $d_{yz}$ orbitals (see supplementary information), transforming as irreducible representations B$_1$ and B$_2$, respectively. The bottommost conduction band at both valleys mainly originates from the $d_{x^2-y^2}$ orbital, with a minor contribution from the $d_{z^2}$ orbital, both of which belong to the A$_1$ representation. In optical transitions, the momentum operator $\langle\psi_{c\mathbf{k}}|\mathbf{p}|\psi_{v\mathbf{k}}\rangle$ transforms as a polar vector, decomposing into B$_1$ and B$_2$ representations at the X and Y valleys, respectively. According to the great orthogonality theorem, the dipole matrix elements satisfy the selection rules: $B_1 \otimes B_1 = A_1$, $B_2 \otimes B_2 = A_1$. These results indicate that the *x*-polarized photons selectively excite spin-up electrons at the X valley, whereas *y*-polarized photons excite spin-down electrons at the Y valley, as illustrated in Fig. 2(d). Because of the same symmetry, these optical selection rules are expected to apply to other Lieb-lattice altermagnets, e.g. M$_2$WX$_4$ (M=Fe, X=S, Se) (see supplementary information).

Next, we investigate the excitonic effects by computing the electron-hole interaction kernel and solving the Bethe-Salpeter Equation (BSE) [48],

$$\left(E_{c\mathbf{k}}^{QP} - E_{v\mathbf{k}}^{QP}\right)A_{vc\mathbf{k}}^{S} + \sum_{v'c'\mathbf{k}'}\langle vc\mathbf{k}|K^{\text{eh}}|v'c'\mathbf{k}'\rangle = \Omega^{S}A_{vc\mathbf{k}}^{S}, \qquad (2)$$

where $A_{vc\mathbf{k}}^S$ is the exciton wavefunction coefficient and $\Omega^S$ is the exciton energy. $K^{eh}$ represents the electron-hole interaction kernel. Using BSE, we have computed the optical absorbance spectrum, defined as $A(\omega) = \frac{\omega d}{c}\varepsilon_2(\omega)$, where $d$ is the interlayer spacing between adjacent Mn$_2$WS$_4$ layers along the out-of-plane direction. $\varepsilon_2(\omega)$ is the imaginary part of the dielectric function, which is expressed as

$$\epsilon_2(\omega) = \frac{16\pi^2 e^2}{\omega^2}\sum_S |\mathbf{e}\cdot\langle 0|\mathbf{v}|S\rangle|^2 \delta(\omega-\Omega^S), \quad (3)$$

where $\langle 0|\mathbf{v}|S\rangle = \sum_{vc\mathbf{k}} A_{vc\mathbf{k}}^S \langle\psi_{v\mathbf{k}}|\mathbf{v}|\psi_{c\mathbf{k}}\rangle$ denotes the transition matrix element from the ground (vacuum) state $|0\rangle$ to an exciton state $|S\rangle$. As shown in Fig. 3(a), under the incident light polarized along the $x$-direction, the solid curve denotes the absorbance spectrum obtained from the BSE, which incorporates electron-hole interactions, while the dashed curve represents the single-particle spectrum without excitonic effects. A prominent excitonic peak, labeled $X_1^x$, appears at 0.52 eV, corresponding to a large exciton binding energy of 1.14 eV. This large binding energy indicates significant Coulomb interactions in the reduced dielectric environment at the 2D limit [56,57]. Meanwhile, we have plotted the absorbance spectrum under $y$-polarized excitation ($\epsilon_y$) in Fig. 3(b). It is identical to that under $x$-polarization ($\epsilon_x$) (Fig. 3(a)), and the optical absorption are also dominated by a strongly bound exciton ($X_1^y$) at the same energy.

We have observed an excitonic valley-select linear dichroism, which has been predicted in the model developed by Cao. etc., using symmetry classification combined with BSE. [63] As shown in the inset of Fig. 3(c), the exciton $X_1^x$ is primarily localized around the X valley but is elongated along the X-M ($y$) direction in reciprocal space. This is due to the highly anisotropic band dispersion (effective mass), as depicted in Fig. 1(c). Correspondingly, the real-space wavefunction of exciton $X_1^x$, shown in Fig. 3(c), is highly anisotropic but elongated along the *x*-direction, which is orthogonal to its reciprocal space distribution. This spatial anisotropy aligns with the polarization dependence observed in Fig. 3(a), where the elongation direction coincides with the polarization of the incident light, leading to a large dipole oscillator strength. The strongly anisotropic nature of the wavefunction also implies direction-dependent excitonic dynamics, with enhanced dispersion and mobility along the *x*-direction.

In contrast, the degenerate exciton $X_1^y$ exhibits complementary characteristics. It is mainly located at the Y valley and elongated along the M-Y (*x*-) direction in reciprocal space. Its real-space wavefunction is correspondingly anisotropic and aligned along the *y* direction, consistent with the large oscillator strength observed under *y*-polarized excitation.

Beyond this valley locking effect, the spin is also locked to specific excitons due to the altermagnetic spin structure. Figs. 3(c) and (d) display the dominant spinor components of these two excitons, represented by blue and red color maps, respectively. The minor spin components for each exciton are found to be nearly negligible. This clearly supports that both excitons are spin-polarized with opposite spin orientations. These spin-polarized excitonic states offer a promising stage for encoding, storing, and reading valley information via their intertwined spin and valley degrees of freedom. It is also particularly attractive for applications in spin-valleytronic devices, where selective excitation and manipulation of spin/valley-polarized states are essential.

Furthermore, we have explored exciton spectrum and fine structures. Taking the *x*-polarized exciton spectrum as an example. In addition to the lowest-energy exciton ($X_1^x$), the real-space wavefunction of the higher-energy excitons, $X_2^x$ and $X_3^x$, with binding energies of 0.88 eV and 0.83 eV, respectively, are shown in Fig. 3(e) and 3(f). Along with the lowest-energy exciton $X_1^x$, excitons $X_2^x$ and $X_3^x$ exhibit similarly anisotropic distributions localized along the two horizontal manganese atomic chains. However, a key distinction emerges with higher-energy states: additional nodal structures with the higher quantum number. This is more clearly visualized in reciprocal space. As shown in the insets of Figs. 3(c)-(f), the nodal features evolve along the $y$-direction in reciprocal space, enabling a possible assignment of $X_1^x$, $X_2^x$, and $X_3^x$ to the 1s, 2s, and 2p hydrogen-like excitonic states, respectively.

Notably, similar anisotropic exciton series have been observed in monolayer black phosphorus (BP), where the anisotropy arises from the intrinsic crystalline anisotropy of the puckered lattice. [58] In contrast, monolayer $Mn_2WS_4$ possesses an approximately isotropic in-plane crystal structure; the observed anisotropic excitons originate from the underlying spin splitting associated with the alternating spin texture of the manganese sublattices. More importantly, these excitons are fully spin-polarized. These distinctions underscore the fundamentally different mechanism for exciton anisotropy, driven by altermagnetic ordering rather than lattice geometry.

SVL plays a crucial role in transport and optical properties of TMDs and underpins many of their potential applications. However, due to the degeneracy of the valleys, photoexcited carriers can be relaxed between them, thereby reducing the spin and valley polarization. Intriguingly, the valley degeneracy can be lifted by applying uniaxial strain in $S_4T$-symmetric Lieb-lattice altermagnets. Such symmetry breaking leads to an energy splitting of the conduction and valence band extrema at the X and Y valleys (see supplementary information). As a result, the optical absorbance spectra for $\epsilon_x$ and $\epsilon_y$ component are no longer identical. As shown in Fig. 4(a), under 2% tensile uniaxial strain along the $x$-direction, the $X_1^x$ exciton exhibits a redshift, while the $X_1^y$ exciton remains unaffected compared to the unstrained case, resulting in an emergent energy splitting Δ around 75 meV.

The energy splitting of two excitons serves as a fingerprint of strain-tunable, valley-selective excitonic physics in altermagnets. Fig. 4(b) shows the evolution of the energy splitting $\Delta$ as a function of uniaxial tensile strain applied along the $x$-direction. The red curve includes excitonic effects from the BSE, while the gray curve represents single-particle results without electron-hole interactions, capturing only the underlying band-structure contribution. The enhanced splitting $\Delta$ observed in the excitonic spectrum (red curve) indicates that the strain sensitivity of valley excitons can be amplified by excitonic effects.

This unique strain-induced tunability enables selective control and manipulation of excitons across both spin and valley degrees of freedom. As shown in Fig. 4(c), under uniaxial tensile strain along the $x$-direction, the lowest-energy exciton becomes localized at the X valley. As a result, excited states will be ultimately relaxed to this energetically favorable excitonic state, leading to an ensemble of fully spin-up-polarized excitons. Correspondingly, the resulting photoluminescence (PL) emission is expected to be linearly polarized along the $x$-direction.

The spin polarization and PL of excitons can be efficiently tuned by strain. As shown in Fig. 4(d), when stretching along the $y$ direction, the lowest-energy exciton is switched to the Y valley. Correspondingly, the photo-collected excitons will be fully spin-up polarized, and the PL is linearly polarized along the y direction.

It is worth mentioning that this strain-tunable SVL cannot be realized in TMDs. The three-fold in-plane rotational symmetry of TMDs protects the valley degeneracy from uniaxial strain, necessitating the application of a magnetic field to break that degeneracy. In contrast, the established strain engineering in four-fold altermagnets provides a more practical and versatile means of controlling spin-valley properties compared to applying a magnetic field.

Finally, we anticipate that our observed valley-selective linear dichroism and excitonic effects can be generalized to other four-fold-symmetric altermagnets, such as the A(BN)$_2$ (A = Mg, Ca, Zn and B = Mn, Fe, Co) family [59]. Moreover, future investigations into magneto-optical phenomena, such as Kerr rotation and spin-polarized photocurrent generation, are expected to further illuminate the rich spin-valley and excitonic physics in Lieb-lattice altermagnets. In addition, the twisted moiré altermagnetic systems are also expected to be formed by stacking two monolayer Lieb lattices [60–62], which might provide a promising and highly tunable platform for exploring emergent quantum phenomena arising from the interplay of spin, valley, and moiré degrees of freedom.

**Acknowledgment**

This work is supported by the National Science Foundation DMR- 2124934. The simulation used Anvil at Purdue University through allocation DMR100005 from the Advanced Cyberinfrastructure Coordination Ecosystem: Services & Support (ACCESS) program, which is supported by National Science Foundation grants #2138259, #2138286, #2138307, #2137603, and #2138296.

Figures:

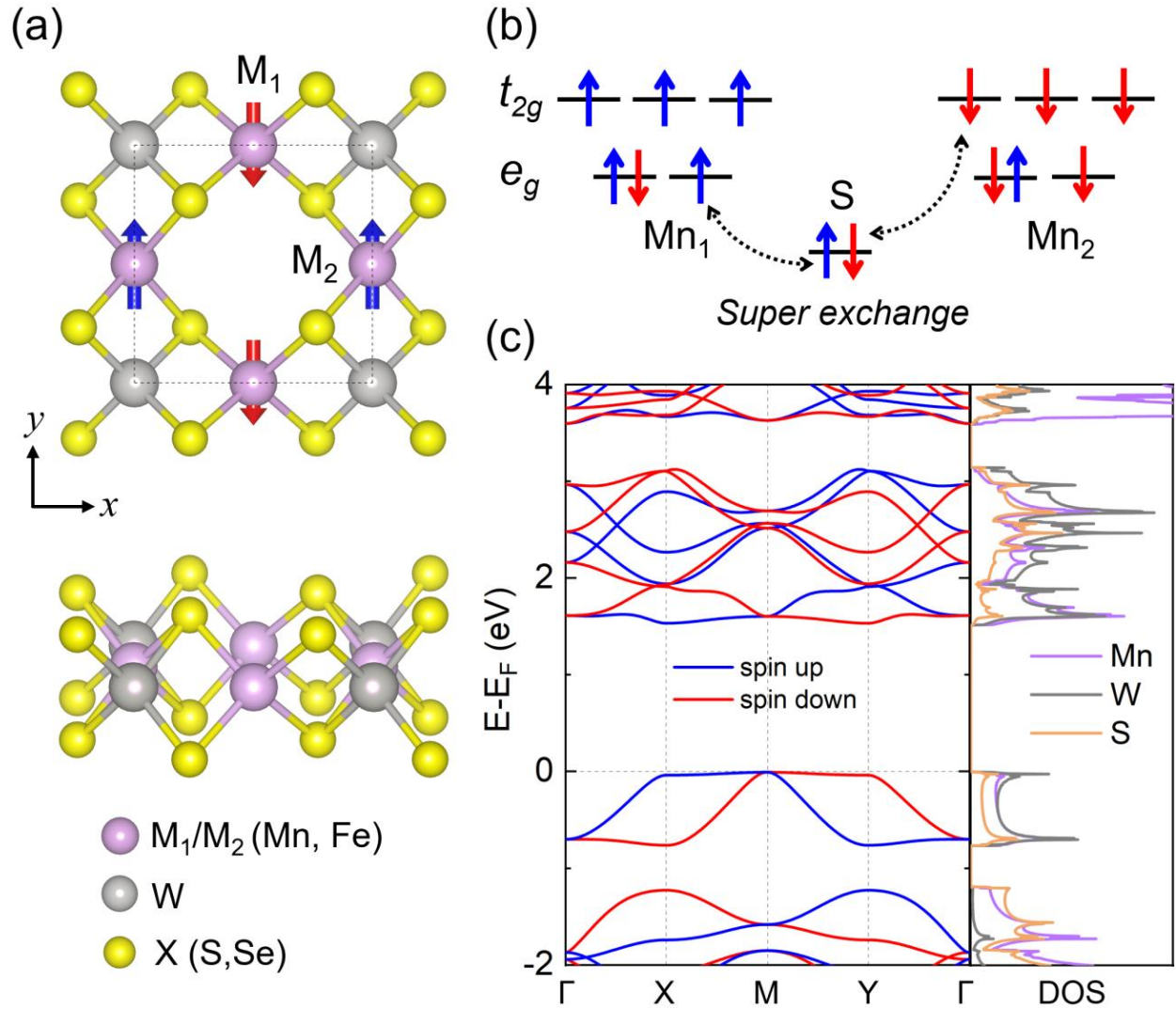

**Fig. 1.** (a) Top and side views of the atomic structure of the Lieb-lattice altermegnets in monolayer $M_2WX_4$ (M = Mn, Fe; X = S, Se). Blue and red arrows indicate the spin-up and spin-down magnetization of the M atoms, respectively. (b) Superexchange interactions between $Mn_1$ and $Mn_2$ atoms mediated by sulfur atoms. (c) Spin-polarized band structure and density of states (DOS) of monolayer $Mn_2WS_4$. Blue and red curves correspond to spin-up and spin-down components, respectively.

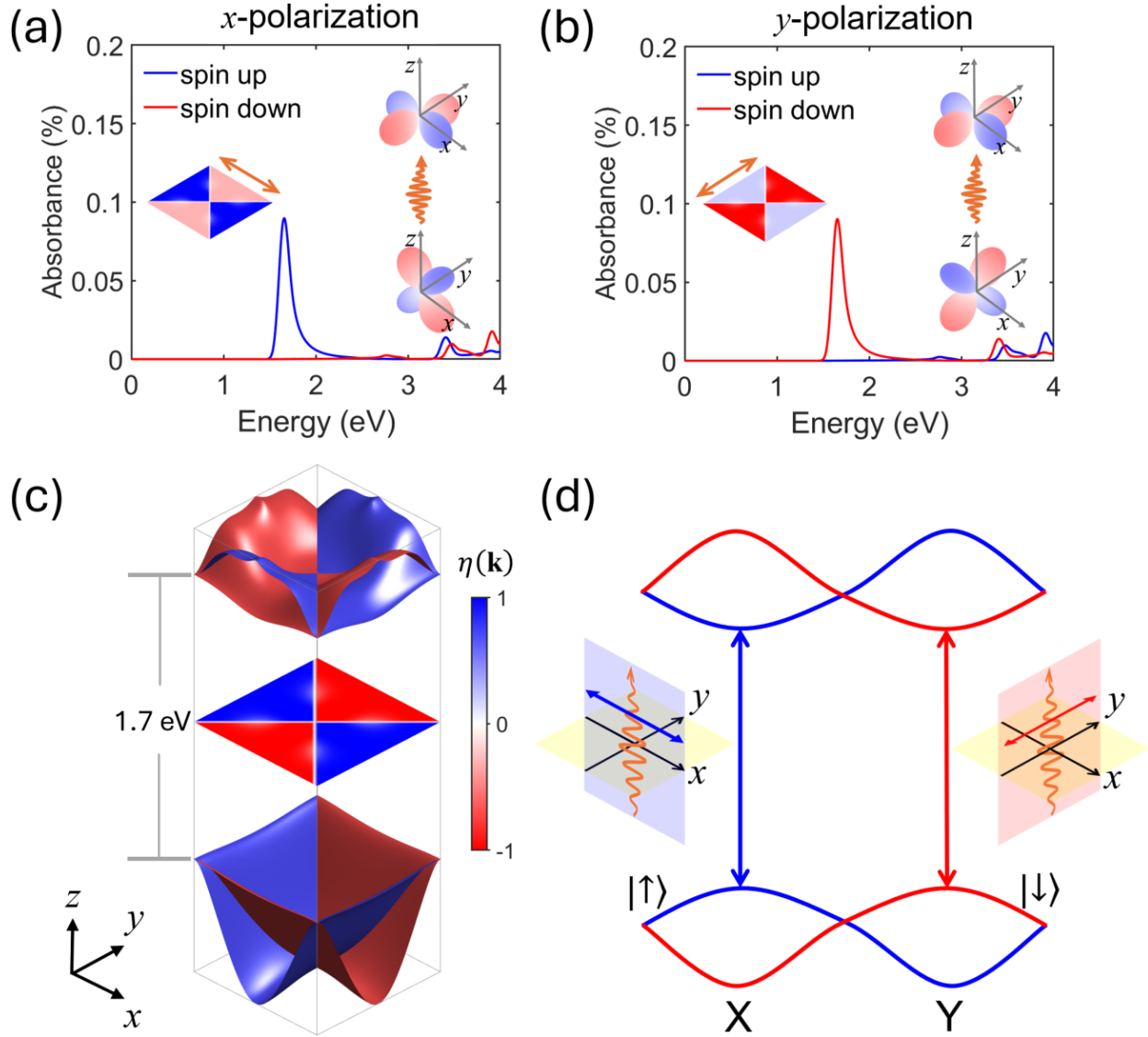

**Fig. 2.** (a–b) Optical absorption spectrum of monolayer $Mn_2WS_4$ under linearly polarized light along the *x*- and *y*-directions, respectively, decomposed into spin-up and spin-down components. Insets depict the atomic orbitals involved in the interband transitions and their momentum-resolved distributions. (c) Three-dimensional band structure (first valence and conduction bands) of monolayer $Mn_2WS_4$. The middle plane represents the Brillouin zone, showing the momentum-resolved degree of linear optical polarization, $\eta(\mathbf{k})$, as defined in the main text. (d) Schematic plot illustrating the valley-dependent optical selection rules. Blue (red) curves represent the bands with spin-up (spin-down) components.

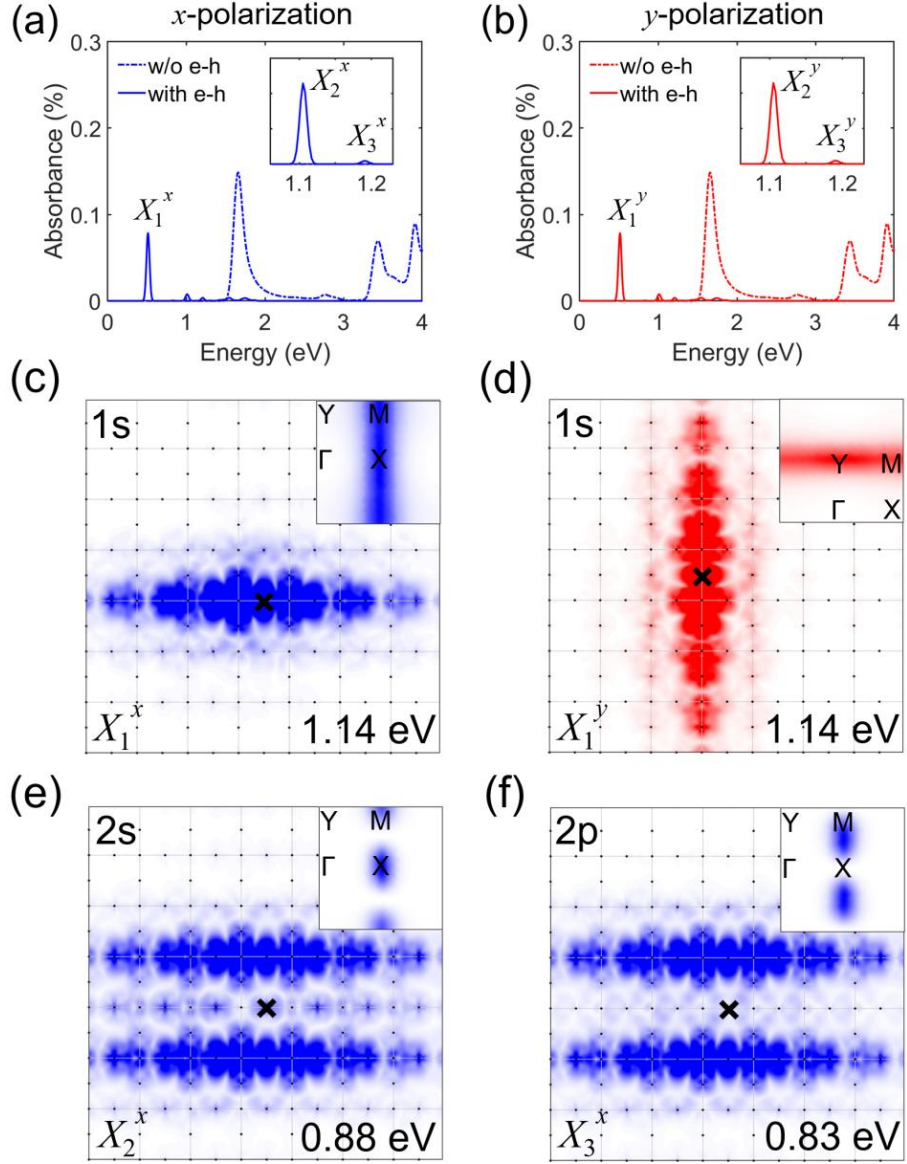

**Fig. 3** (a)-(b) Optical absorbance spectrum calculated using the GW-BSE approach. The solid (dashed) blue and red curve represent the spectrum with (without) electron–hole interactions under $x$- and $y$-polarization, respectively. (c)-(f) Modulus of real-space exciton wavefunctions for the excitons $X_1^x$, $X_1^y$, $X_2^x$, and $X_3^x$. The black cross symbol denotes the fixed hole positions at the manganese sublattices. Insets show the corresponding momentum-resolved exciton wavefunction coefficients $|A_{vc\mathbf{k}}^S|^2$. Blue and red colors represent the spin-up and spin-down channels, respectively.

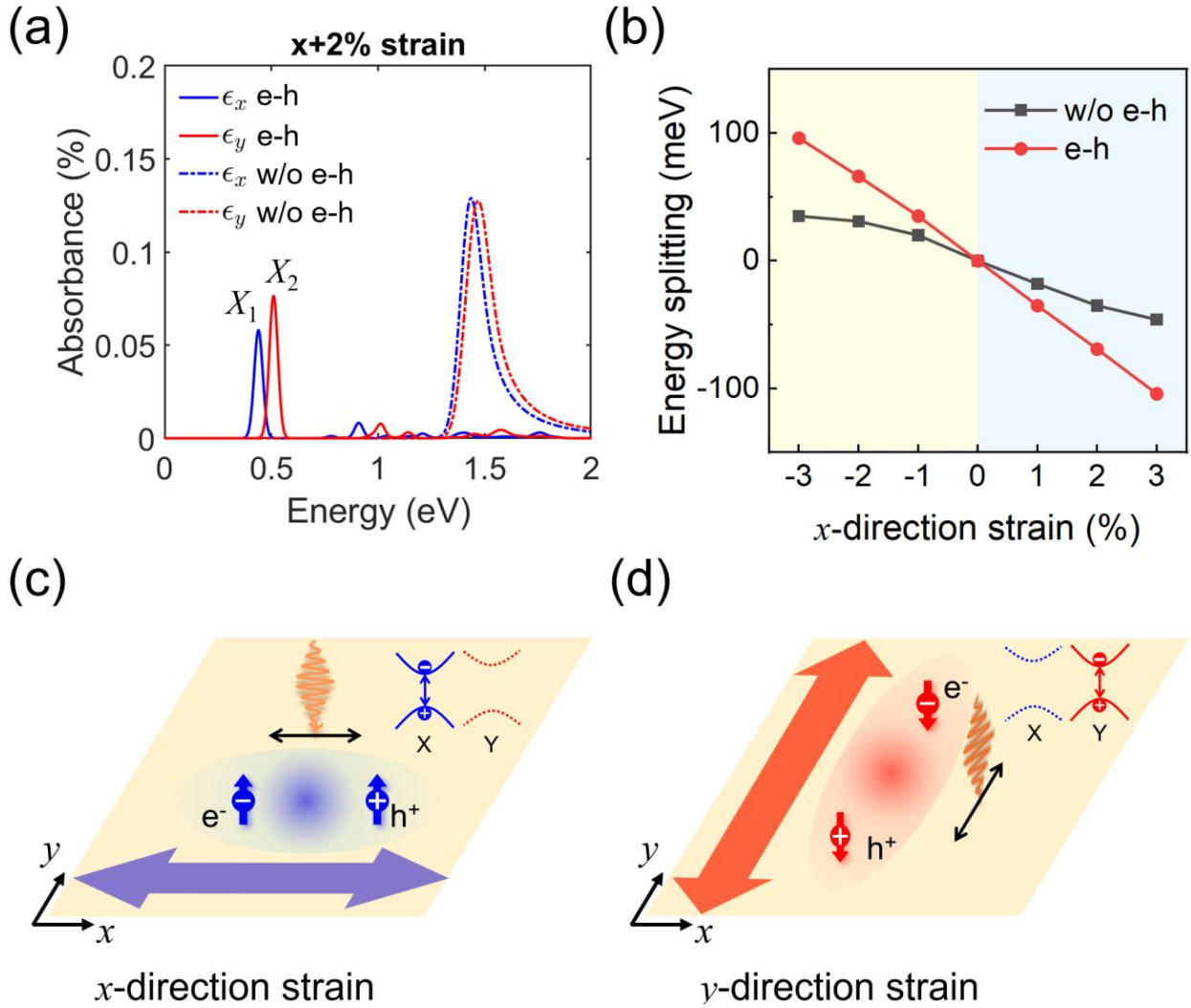

**Fig. 4** (a) Optical absorption spectra $\epsilon_x$ and $\epsilon_y$ for monolayer $Mn_2WS_4$ under 2% uniaxial tensile strain applied along the $x$-direction. (b) Evolution of the energy splitting between the two exciton peaks $X_1^x$ and $X_1^y$ as a function of uniaxial strain along the $x$-direction. The red curve includes electron-hole interactions, while the gray curve represents the result without considering electron-hole interactions. (c)-(d) Schematic illustration of spin-polarized excitons in monolayer $Mn_2WS_4$ under uniaxial strain along the $x$- and $y$-directions, respectively.